# DNA evolved to minimize frameshift mutations

Valentina Agoni valentina.agoni@unipv.it

**Point mutations can surely be dangerous but what is worst than to lose the reading frame?! Does DNA evolved a strategy to try to limit frameshift mutations?! Here we investigate if DNA sequences effectively evolved a system to minimize frameshift mutations analyzing the transcripts of proteins with high molecular weights.**

During replication the DNA polymerase creates an error every $10^5$-$10^6$ nucleotides [1] [2]. In [3] we demonstrate that the amino acids coding table evolved to minimize point mutations. They can of course be dangerous but what is worst than to lose the reading frame?! So how to limit frameshift mutations? To verify if the nucleotide sequences (not the amino acids coding table as for point mutations [3]) evolved to minimize the probability of frameshift mutations we analized the amino acids coding table to find the proper sequences to consider.

We are interested in a sequence the more homogeneous as possible in terms of nucleotides composition to compare its frequency with a less homogeneous one. First there are of course 4 codons composed by the repetition of the same nucleotide: AAA, CCC, GGG and TTT (blue circles in Figure 1). If we look at the amino acids coding table we notice that 3 of them code for amino acids that are associated also with many other triplets. On the contrary we want to compare the frequency of 2 sequences undergoing differences due to the fact that the associated amino acids could be coded by a diverse number of codons.

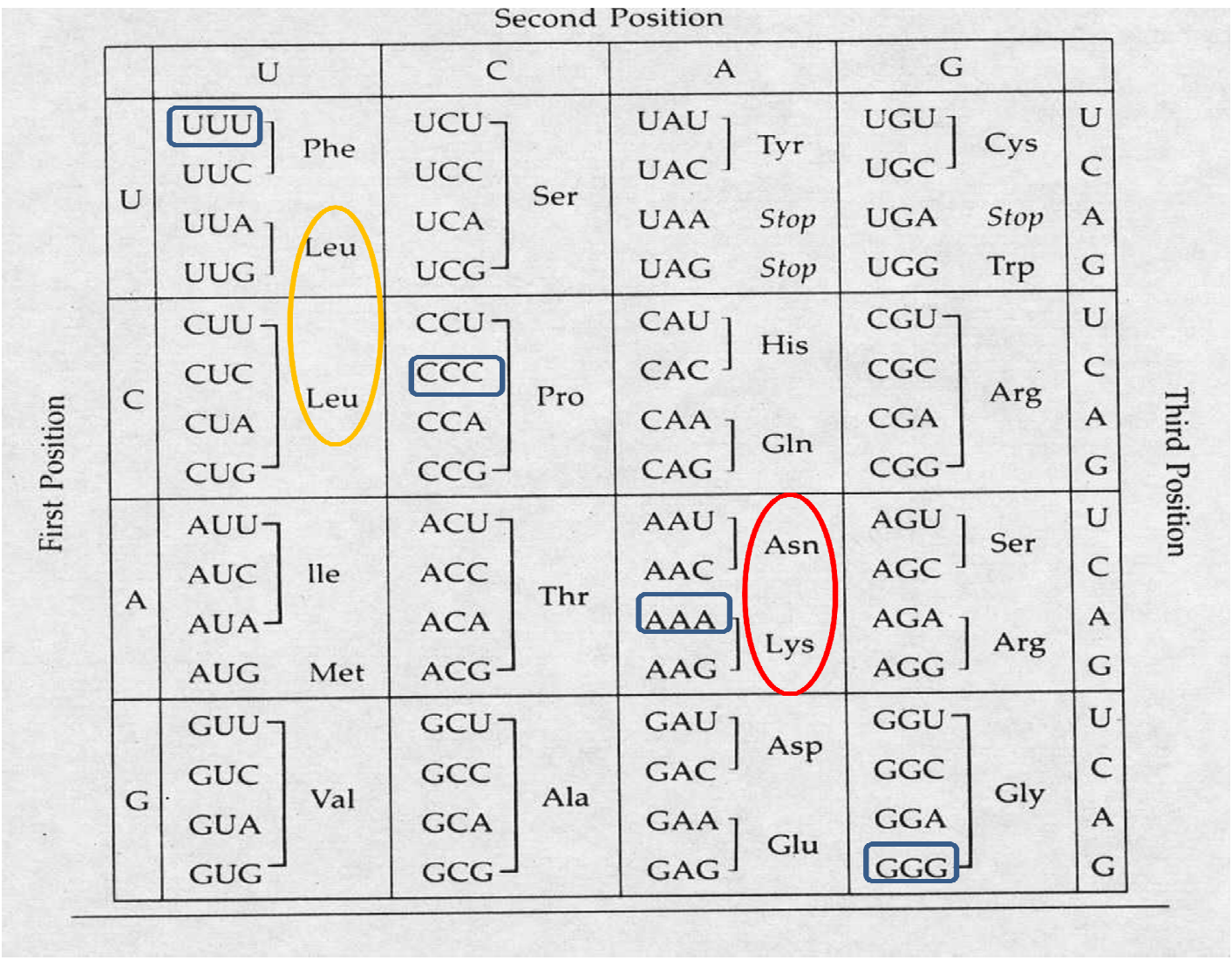

| First Position | U | C | A | G | Third Position |
|---|---|---|---|---|---|
| U | UUU Phe<br>UUC Phe<br>UUA Leu<br>UUG Leu | UCU Ser<br>UCC Ser<br>UCA Ser<br>UCG Ser | UAU Tyr<br>UAC Tyr<br>UAA Stop<br>UAG Stop | UGU Cys<br>UGC Cys<br>UGA Stop<br>UGG Trp | U<br>C<br>A<br>G |
| C | CUU Leu<br>CUC Leu<br>CUA Leu<br>CUG Leu | CCU Pro<br>CCC Pro<br>CCA Pro<br>CCG Pro | CAU His<br>CAC His<br>CAA Gln<br>CAG Gln | CGU Arg<br>CGC Arg<br>CGA Arg<br>CGG Arg | U<br>C<br>A<br>G |
| A | AUU Ile<br>AUC Ile<br>AUA Ile<br>AUG Met | ACU Thr<br>ACC Thr<br>ACA Thr<br>ACG Thr | AAU Asn<br>AAC Asn<br>AAA Lys<br>AAG Lys | AGU Ser<br>AGC Ser<br>AGA Arg<br>AGG Arg | U<br>C<br>A<br>G |
| G | GUU Val<br>GUC Val<br>GUA Val<br>GUG Val | GCU Ala<br>GCC Ala<br>GCA Ala<br>GCG Ala | GAU Asp<br>GAC Asp<br>GAA Glu<br>GAG Glu | GGU Gly<br>GGC Gly<br>GGA Gly<br>GGG Gly | U<br>C<br>A<br>G |

**Figure 1. The amino acids coding table.** We are interested in finding 2 codons constituted by the repetition of the same nucleotide for the higher number times. The amino acids coding table reveals that for our purpose we cannot use Phe and Leu because it can be coded by many codons. On the other hand the Lys-Arg duo is perfect.

We cannot compare for example the frequency of AAA with the frequency of GGG because they are affected for example by the amino acids frequency and by the number of codons associated with a single amino acid. Therefore we focalized on Lys (K) that can be codified by AAA or AAG. Moreover a single codon is not representative to check the frameshift probability because we know that a shift of DNA polymerase is much higher the longer is the sequence composed by the repetition of a single nucleotide. The probability of a pre-extablished amino acids triplet was too low even in long transcripts.

In particular we select the amino acids pair LysArg (KN) because:

- there is only one other amino acids codified only by a similar sequence of nucleotides: Asn (N) that can be codified by AAU or AAC
- both these 2 amino acids could be coded only by the 2 codons under investigation
- moreover to study an amino acids couple allows us to eliminate problems related to their different frequency in the DNA.

In this way the comparison will be balanced. Hence the best thing to do is to consider the couple KN namely to check the frequency of AAGAAC and AAGAAT respect to the frequency of AAAAAC and AAAAAT.

We analyzed proteins with a high molecular weight (that means with long transcripts) in order to have a great number of KN. This number is obviously influenced by Single Nucleotide Polymorphisms (SNPs) and by alternative splicing. It is useful to clarify that we analyzed the frequency of 2 amino acids and into mRNAs despite the principle of frameshift mutations minimization is relative to DNA replication for the reasons listed before. If the nucleotide sequences evolved to minimize the probability of frameshift mutations we should find AAGAAC and AAGAAT (green in Figure 2) more than AAAAAC and AAAAAT (yellow). This is what we actually found. Figure 2 reports the sequences analyzed: human myosin [4], human BRCA1 (breast cancer type 1)[5] and murine 53BP1 [6]. We evaluated the ratio between green sequences and green plus yellow sequences. The results are respectively: 1, 0.7 and 0.6 with an average of 0.77. Comparing it with the equal probability to have green and yellow sequences (ratio: 0.5 with a negligible standard deviation) using a one tailed t-Student's test we have a p value of 0.04537.

# H.sapiens mRNA for myosin

GenBank: Z38133.1

```
>gi|558668|emb|Z38133.1| H.sapiens mRNA for myosin
ATGAGTGCGAGCTCAGACGCTGAGATGGCTGTTTTTGGCGAACGTGCTCCCTACCTTCGAAAATCA
GAAAAGGAGCGGATTGAGGCCCAAAACAAGCCGTTTGATGCTAAAACATCTGTCTTTGTGGCGGAG
CCCAAGGAATCCTATGTGAAGAGCACTATACAAAGCAAAGAAGGAGGGAAAGTAACCGTAAAGACT
GAAGGTGGAGCAACTCTAACTGTCAGGGAAGACCAAGTCTTCCCTATGAACCCTCCGAAATATGAC
AAAATTGAGGACATGGCCATGATGACTCATCTACACGAGCCTGGAGTGCTGTACAACCTCAAAGAG
CGCTATGCAGCCTGGATGATCTACACCTACTCAGGCCTCTTCTGTGTCACCGTCAACCCCTACAAG
TGGCTGCCGGTGTACAAGCCCGAGGTGGTGGCTGCCTACAGAGGCAAAAAGCGCCAGGAGGCCCCG
CCCCACATCTTCTCCATCTCTGACAATGCCTATCAGTTCATGTTGACTGATCGAGAGAATCAGTCC
ATCCTGATCACCGGAGAATCTGGTGCCGGAAAGACTGTGAACACCAAGCGTGTCATCCAATACTTT
GCAACAATTGCAGTTACTGGAGAGAAGAAGAAGGATGAATCTGGCAAAATGCAGGGGACTCTGGAA
GATCAAATCATCAGCGCCAATCCCCTACTGGAGGCCTTTGGCAATGCCAAAACCGTGAGGAATGAC
AACTCCTCTCGCTTTGGTAAATTCATTAGAATCCACTTTGGTACTACAGGGAAGCTGGCATCTGCT
GATATAGAAACATATCTTTTAGAAAAGTCCAGAGTTACTTTCCAGCTAAAGGCGGAAAGAAGCTAC
CATATTTTTTATCAGATCACTTCCAATAAGAAGCCAGATCTAATTGAAATGCTCCTGATCACCACC
AACCCATATGACTATGCCTTCGTCAGTCAGGGGGAGATCACAGTTCCCAGTATTGATGACCAAGAA
GAGTTGATGGCCACTGATAGTGCCATTGACATCCTGGGCTTCACTCCTGAAGAGAAAGTGTCCATC
TATAAACTCACAGGGGCTGTGATGCATTATGGGAACATGAAATTCAAGCAAAAGCAGCGTGAGGAG
CAAGCTGAGCCAGATGGCACAGAAGTCGCTGACAAGGCAGCCTATCTCCAGAGTCTGAACTCTGCA
GACCTACTCAAAGCCCTCTGCTACCCTAGGGTCAAGGTTGGCAATGAGTATGTCACCAAAGGCCAG
ACTGTGCAGCAGGTGTACAATGCGGTGGGTGCTCTGGCCAAAGCCGTCTACGAGAAGATGTTCCTG
TGGATGGTCACCCGCATCAACCAGCAGCTGGACACCAAGCAGCCCAGGCAGTACTTCATCGGGGTC
TTGGACATTGCTGGCTTTGAAATCTTTGATTTTAACAGCCTGGAGCAGCTGTGCATCAACTTCACC
AACGAGAAACTGCAACAGTTTTTCAACCACCACATGTTTGTGCTAGAGCAGGAGGAGTACAAGAAG
GAAGGCATCGAGTGGACGTTCATTGACTTTGGGATGGACCTGGCTGCCTGCATTGAGCTCATTGAG
AAGCCACTGGGCATCTTCTCCATCCTGGAAGAGGAGTGCATGTTCCCTAAGGCAACGGACACCTCC
TTCAAGAACAAGCTGTATGACCAGCACCTGGGCAAGTCTGCCAACTTCCAGAAGCCCAAGGTGGTC
AAAGGCAAGGCTGAGGCCCACTTCTCTCTGATTCACTATGCTGGCACTGTGGACTACAACATTACT
GGCTGGCTGGACAAAAATAAGGACCCCCTGAATGATACTGTGGTTGGGCTGTACCAGAAGTCTGCA
ATGAAGACTCTAGCCAGTCTCTTTTCCACGTATGCTAGTGCTGAAGCAGATAGCAGCGCGAAGAAA
GGTGCTAAGAAAAAGGGCTCTTCTTTCCAGACTGTGTCTGCCCTTTTCAGGGAAAATTTAAATAAA
TTGATGACGAATCTGAGGAGCACACACCCTCACTTCGTACGGTGTATCATTCCCAATGAAACCAAA
ACTCCTGGGGCAATGGAACATGAACTTGTGTTGCACCAGCTGAGGTGTAATGGTGTGCTGGAAGGC
ATCCGCATCTGTAGGAAAGGATTCCCAAGCAGAATCTTATATGGTGATTTCAAACAAAGATACAAG
GTTTTAAATGCAAGTGCTATTCCAGAGGGACAGTTCATTGACAGCAAGAAGGCTTCTGAGAAACTT
CTTGCATCTATTGATATTGATCATACTCAATATAAATTTGGACATACCAAGGTTTTCTTCAAAGCT
GGACTTCTGGGTCTTCTGGAAGAAATGAGAGATGAAAAATTAGCCCAAATTATAACAAGAACACAA
GCTGTCTGTAGGGGATTCCTAATGAGGGTAGAATATCAGAAGATGTTGCAAAGGAGAGAAGCACTT
TTCTGCATCCAGTATAATGTCCGTGCCTTCATGAACGTCAAGCACTGGCCCTGGATGAAACTCTTT
TTCAAGATTAAGCCCCTCCTCAAGAGTGCAGAGACCGAGAAAGAGATGGCCACCATGAAGGAAGAA
TTCCAGAAAACCAAAGATGAACTCGCCAAGTCAGAGGCAAAACGGAAGGAGCTAGAGGAAAAAATG
GTCACTCTCTTAAAAGAGAAAAATGACCTGCAACTCCAGGTTCAATCTGAAGCAGATAGCTTGGCT
GATGCAGAGGAAAGGTGTGAGCAACTGATTAAAAACAAAATCCAACTTGAGGCCAAAATCAAAGAG
```

GTGACTGAAAGAGCTGAGGAGGAGGAAGAGATCAATGCTGAGCTGACAGCCAAGAAGAGAAAACTG

GAGGATGAATGTTCAGAACTCAAGAAAGACATTGATGACCTTGAGCTGACACTGGCCAAGGTTGAG

AAGGAGAAACATGCCACGGAGAACAAGGTGAAAAATCTTACAGAAGAGATGGCAGGCCTGGATGAA

ACCATTGCAAAACTGTCCAAGGAGAAGAAGGCTCTCCAAGAGACCCACCAGCAGACCCTGGATGAC

CTGCAGGCAGAGGAGGACAAAGTCAACATCCTGACCAAAGCTAAAACCAAGCTAGAACAGCAAGTG

GATGATCTTGAAGGGTCTCTGGAACAAGAAAAGAAGCTTCGAATGGATCTAGAAAGAGCAAAGCGG

AAACTGGAGGGTGACCTCAAATTGGCCCAAGAATCCACAATGGATATGGAAAATGACAAACAGCAA

CTTGATGAAAAGCTTGAAAAGAAAGAATTTGAAATCAGCAATTTGATAAGCAAAATTGAAGATGAG

CAAGCTGTAGAAATTCAACTACAGAAGAAGATCAAAGAGTTGCAGGCCCGCATTGAGGAGCTGGGG

GAAGAAATCGAGGCAGAGAGGGCGTCCCGAGCCAAAGCGGAGAAGCAGCGCTCTGACCTCTCCCGG

GAACTGGAGGAGATCAGCGAGAGGCTGGAAGAAGCCGGTGGGGCAACTTCTGCTCAGGTGGAATTG

AACAAGAAGCGGGAGGCTGAGTTTCAGAAACTGCGCAGGGACCTGGAGGAGGCCACCCTGCAGCAT

GAAGCTATGGTGGCTGCTCTTCGGAAGAAGCACGCAGACAGTATGGCTGAGCTTGGGGAGCAGATT

GACAACTTGCAGCGGGTCAAACAGAAGCTGGAGAAGGAGAAGAGTGAGCTGAAGATGGAGACTGAT

GACCTCAGCAGTAACGCAGAGGCCATTTCCAAAGCCAAGGGAAACCTTGAAAAGATGTGCCGCTCT

CTAGAAGATCAAGTGAGTGAGCTTAAGACCAAGGAAGAGGAGCAGCAGCGGCTGATCAATGACCTC

ACAGCACAGAGAGCGCGCCTGCAGACAGAAGCGGGTGAATATTCTCGACAATTAGATGAGAAAGAT

GCTTTAGTCTCTCAGCTTTCAAGGAGCAAGCAAGCATCTACTCAGCAGATTGAAGAGCTGAAACAT

CAACTAGAGGAAGAAACTAAAGCCAAGAACGCCCTGGCACACGCCCTGCAGTCCTCCCGCCATGAC

TGCGACCTGCTGCGGGAACAGTATGAGGAAGAGCAGGAAGGCAAAGCTGAGCTGCAGAGGGCGCTG

TCCAAGGCCAACAGTGAGGTTGCCCAGTGGAGAACCAAATACGAGACGGATGCCATCCAGCGCACA

GAGGAGCTGGAGGAGGCCAAGAAAAAGTTGGCCCAGCGCCTGCAAGAAGCTGAGGAACATGTAGAA

GCTGTGAACGCCAAATGTGCTTCCCTTGAGAAGACGAAGCAGCGGCTCCAGAATGAAGTTGAAGAC

CTCATGCTTGATGTGGAAAGGTCTAATGCAGCCTGTGCAGCCCTTGATAAGAAGCAAAGGAACTTT

GACAAGGTCCTATCAGAATGGAAGCAGAAGTATGAGGAAACTCAGGCTGAACTTGAGGCCTCCCAG

AAGGAGTCACGTTCTCTTAGCACTGAGCTGTTCAAGGTGAAGAATGTCTATGAGGAATCCCTGGAT

CAACTCGAAACGCTAAGAAGAGAAAATAAGAACTTGCAACAGGAGATTTCTGACCTCACTGAGCAG

ATTGCAGAGGGAGGAAAGCAAATTCATGAATTGGAGAAAATAAAGAAGCAAGTAGAACAAGAGAAA

TGTGAAATTCAGGCTGCTTTAGAGGAAGCAGAGGCATCTCTTGAACATGAAGAAGGAAAGATTCTG

CGTATCCAGCTTGAGTTAAACCAAGTCAAGTCTGAAGTTGATAGAAAAATCGCAGAAAAGGATGAG

GAAATTGACCAGCTGAAGAGAAACCACACTAGAGTCGTGGAGACAATGCAGAGCACGCTGGATGCA

GAGATTAGAAGCAGAAATGATGCTCTGAGAGTCAAGAAGAAAATGGAAGGAGATCTGAATGAAATG

GAAATCCAGCTGAACCATGCCAATCGCTTAGCTGCAGAGAGTTTAAGGAACTACAGGAACACCCAA

GGAATCCTGAAGGAAACCCAGCTCCACCTGGATGATGCTCTCCGGGGCCAGGAGGACCTCAAGGAA

CAGCTGGCAATTGTGGAGCGCAGAGCCAACCTGCTGCAGGCTGAGATCGAGGAGCTGTGGGCCACT

CTGGAACAGACAGAGAGAAGCAGGAAAATCGCCGAACAGGAGCTCCTGGATGCCAGTGAGCGTGTC

CAGCTCCTCCACACCCAGAATACCAGTCTCATTAACACCAAGAAGAAATTAGAAAATGACGTTTCC

CAACTCCAAAGTGAAGTGGAAGAAGTAATCCAAGAATCACGCAATGCAGAAGAGAAAGCCAAGAAG

GCCATCACTGATGCTGCCATGATGGCTGAGGAGCTGAAGAAGGAACAGGACACCAGCGCCCACCTG

GAGCGGATGAAGAAGAACCTGGAGCAGACGGTGAAGGACCTGCAGCATCGTCTAGATGAGGCCGAG

CAGCTGGCGCTGAAGGGTGGGAAGAAGCAGATCCAGAAACTGGAGGCCAGGGTACGTGAGCTTGAA

GGAGAGGTTGAAAATGAACAGAAACGTAATGCAGAGGCTGTTAAAGGTTTACGGAAACATGAGCGA

CGAGTAAAAGAACTCACCTACCAGACTGAAGAAGATCGCAAGAATGTTCTCAGGCTGCAGGACTTG[..]

# Homo sapiens breast cancer 1, early onset (BRCA1), transcript variant 1, mRNA

NCBI Reference Sequence: NM_007294.3

```
>gi|237757283|ref|NM_007294.3| Homo sapiens breast cancer 1, early onset (BRCA1), transcript variant 1, mRNA
ATGGATTTATCTGCTCTTCGCGTTGAAGAAGTACAAAATGTCATTAATGCTATGCAGAAAATCTTAGAGTGTCCCATCTGTCTGGAGTTGATCAAGGAACCTGTCTCCACAAAG
TGTGACCACATATTTTGCAAATTTTGCATGCTGAAACTTCTCAACCAGAAGAAAGGGCCTTCACAGTGTCCTTTATGTAAGAATGATATAACCAAAAGGAGCCTACAAGAAAGT
ACGAGATTTAGTCAACTTGTTGAAGAGCTATTGAAAATCATTTGTGCTTTTCAGCTTGACACAGGTTTGGAGTATGCAAACAGCTATAATTTTGCAAAAAAGGAAAATAACTCT
CCTGAACATCTAAAAGATGAAGTTTCTATCATCCAAAGTATGGGCTACAGAAACCGTGCCAAAAGACTTCTACAGAGTGAACCCGAAAATCCTTCCTTGCAGGAAACCAGTCTC
AGTGTCCAACTCTCTAACCTTGGAACTGTGAGAACTCTGAGGACAAAGCAGCGGATACAACCTCAAAAGACGTCTGTCTACATTGAATTGGGATCTGATTCTTCTGAAGATACC
GTTAATAAGGCAACTTATTGCAGTGTGGGAGATCAAGAATTGTTACAAATCACCCCTCAAGGAACCAGGGATGAAATCAGTTTGGATTCTGCAAAAAAGGCTGCTTGTGAATTT
TCTGAGACGGATGTAACAAATACTGAACATCATCAACCCAGTAATAATGATTTGAACACCACTGAGAAGCGTGCAGCTGAGAGGCATCCAGAAAAGTATCAGGGTAGTTCTGTT
TCAAACTTGCATGTGGAGCCATGTGGCACAAATACTCATGCCAGCTCATTACAGCATGAGAACAGCAGTTTATTACTCACTAAAGACAGAATGAATGTAGAAAAGGCTGAATTC
TGTAATAAAAGCAAACAGCCTGGCTTAGCAAGGAGCCAACATAACAGATGGGCTGGAAGTAAGGAAACATGTAATGATAGGCGGACTCCCAGCACAGAAAAAAAGGTAGATCTG
AATGCTGATCCCCTGTGTGAGAGAAAAGAATGGAATAAGCAGAAACTGCCATGCTCAGAGAATCCTAGAGATACTGAAGATGTTCCTTGGATAACACTAAATAGCAGCATTCAG
AAAGTTAATGAGTGGTTTTCCAGAAGTGATGAACTGTTAGGTTCTGATGACTCACATGATGGGGAGTCTGAATCAAATGCCAAAGTAGCTGATGTATTGGACGTTCTAAATGAG
GTAGATGAATATTCTGGTTCTTCAGAGAAAATAGACTTACTGGCCAGTGATCCTCATGAGGCTTTAATATGTAAAAGTGAAAGAGTTCACTCCAAATCAGTAGAGAGTAATATT
GAAGACAAAATATTTGGGAAAACCTATCGGAAGAAGGCAAGCCTCCCCAACTTAAGCCATGTAACTGAAAATCTAATTATAGGAGCATTTGTTACTGAGCCACAGATAATACAA
GAGCGTCCCCTCACAAATAAATTAAAGCGTAAAAGGAGACCTACATCAGGCCTTCATCCTGAGGATTTTATCAAGAAAGCAGATTTGGCAGTTCAAAAGACTCCTGAAATGATA
AATCAGGGAACTAACCAAACGGAGCAGAATGGTCAAGTGATGAATATTACTAATAGTGGTCATGAGAATAAAACAAAAGGTGATTCTATTCAGAATGAGAAAAATCCTAACCCA
ATAGAATCACTCGAAAAAGAATCTGCTTTCAAAACGAAAGCTGAACCTATAAGCAGCAGTATAAGCAATATGGAACTCGAATTAAATATCCACAATTCAAAAGCACCTAAAAAG
AATAGGCTGAGGAGGAAGTCTTCTACCAGGCATATTCATGCGCTTGAACTAGTAGTCAGTAGAAATCTAAGCCCACCTAATTGTACTGAATTGCAAATTGATAGTTGTTCTAGC
AGTGAAGAGATAAAGAAAAAAAAGTACAACCAAATGCCAGTCAGGCACAGCAGAAACCTACAACTCATGGAAGGTAAAGAACCTGCAACTGGAGCCAAGAAGAGTAACAAGCCA
AATGAACAGACAAGTAAAAGACATGACAGCGATACTTTCCCAGAGCTGAAGTTAACAAATGCACCTGGTTCTTTTACTAAGTGTTCAAATACCAGTGAACTTAAAGAATTTGTC
AATCCTAGCCTTCCAAGAGAAGAAAAAGAAGAGAAACTAGAAACAGTTAAAGTGTCTAATAATGCTGAAGACCCCAAAGATCTCATGTTAAGTGGAGAAAGGGTTTTGCAAACT
GAAAGATCTGTAGAGAGTAGCAGTATTTCATTGGTACCTGGTACTGATTATGGCACTCAGGAAAGTATCTCGTTACTGGAAGTTAGCACTCTAGGGAAGGCAAAAACAGAACCA
AATAAATGTGTGAGTCAGTGTGCAGCATTTGAAAACCCCAAGGGACTAATTCATGGTTGTTCCAAAGATAATAGAAATGACACAGAAGGCTTTAAGTATCCATTGGGACATGAA
GTTAACCACAGTCGGGAAACAAGCATAGAAATGGAAGAAAGTGAACTTGATGCTCAGTATTTGCAGAATACATTCAAGGTTTCAAAGCGCCAGTCATTTGCTCCGTTTTCAAAT
CCAGGAAATGCAGAAGAGGAATGTGCAACATTCTCTGCCCACTCTGGGTCCTTAAAGAAACAAAGTCCAAAAGTCACTTTTGAATGTGAACAAAAGGAAGAAAATCAAGGAAAG
AATGAGTCTAATATCAAGCCTGTACAGACAGTTAATATCACTGCAGGCTTTCCTGTGGTTGGTCAGAAAGATAAGCCAGTTGATAATGCCAAATGTAGTATCAAAGGAGGCTCT
AGGTTTTGTCTATCATCTCAGTTCAGAGGCAACGAAACTGGACTCATTACTCCAAATAAACATGGACTTTTACAAAACCCATATCGTATACCACCACTTTTTCCCATCAAGTCA
TTTGTTAAAACTAAATGTAAGAAAAATCTGCTAGAGGAAAACTTTGAGGAACATTCAATGTCACCTGAAAGAGAAATGGGAAATGAGAACATTCCAAGTACAGTGAGCACAATT
AGCCGTAATAACATTAGAGAAAATGTTTTTAAAGAAGCCAGCTCAAGCAATATTAATGAAGTAGGTTCCAGTACTAATGAAGTGGGCTCCAGTATTAATGAAATAGGTTCCAGT
GATGAAAACATTCAAGCAGAACTAGGTAGAAACAGAGGGCCAAAATTGAATGCTATGCTTAGATTAGGGGTTTTGCAACCTGAGGTCTATAAACAAAGTCTTCCTGGAAGTAAT
TGTAAGCATCCTGAAATAAAAAAGCAAGAATATGAAGAAGTAGTTCAGACTGTTAATACAGATTTCTCTCCATATCTGATTTCAGATAACTTAGAACAGCCTATGGGAAGTAGT
CATGCATCTCAGGTTTGTTCTGAGACACCTGATGACCTGTTAGATGATGGTGAAATAAAGGAAGATACTAGTTTTGCTGAAAATGACATTAAGGAAAGTTCTGCTGTTTTTAGC
AAAAGCGTCCAGAAAGGAGAGCTTAGCAGGAGTCCTAGCCCTTTCACCCATACACATTTGGCTCAGGGTTACCGAAGAGGGGCCAAGAAATTAGAGTCCTCAGAAGAGAACTTA
TCTAGTGAGGATGAAGAGCTTCCCTGCTTCCAACACTTGTTATTTGGTAAAGTAAACAATATACCTTCTCAGTCTACTAGGCATAGCACCGTTGCTACCGAGTGTCTGTCTAAG
AACACAGAGGAGAATTTATTATCATTGAAGAATAGCTTAAATGACTGCAGTAACCAGGTAATATTGGCAAAGGCATCTCAGGAACATCACCTTAGTGAGGAAACAAAATGTTCT
GCTAGCTTGTTTTCTTCACAGTGCAGTGAATTGGAAGACTTGACTGCAAATACAAACACCCAGGATCCTTTCTTGATTGGTTCTTCCAAACAAATGAGGCATCAGTCTGAAAGC
CAGGGAGTTGGTCTGAGTGACAAGGAATTGGTTTCAGATGATGAAGAAAGAGGAACGGGCTTGGAAGAAAATAATCAAGAAGAGCAAAGCATGGATTCAAACTTAGGTGAAGCA
GCATCTGGGTGTGAGAGTGAAACAAGCGTCTCTGAAGACTGCTCAGGGCTATCCTCTCAGAGTGACATTTTAACCACTCAGCAGAGGGATACCATGCAACATAACCTGATAAAG
CTCCAGCAGGAAATGGCTGAACTAGAAGCTGTGTTAGAACAGCATGGGAGCCAGCCTTCTAACAGCTACCCTTCCATCATAAGTGACTCTTCTGCCCTTGAGGACCTGCGAAAT
CCAGAACAAAGCACATCAGAAAAAGCAGTATTAACTTCACAGAAAAGTAGTGAATACCCTATAAGCCAGAATCCAGAAGGCCTTTCTGCTGACAAGTTTGAGGTGTCTGCAGAT
AGTTCTACCAGTAAAAATAAAGAACCAGGAGTGGAAAGGTCATCCCCTTCTAAATGCCCATCATTAGATGATAGGTGGTACATGCACAGTTGCTCTGGGAGTCTTCAGAATAGA
AACTACCCATCTCAAGAGGAGCTCATTAAGGTTGTTGATGTGGAGGAGCAACAGCTGGAAGAGTCTGGGCCACACGATTTGACGGAAACATCTTACTTGCCAAGGCAAGATCTA
GAGGGAACCCCTTACCTGGAATCTGGAATCAGCCTCTTCTCTGATGACCCTGAATCTGATCCTTCTGAAGACAGAGCCCCAGAGTCAGCTCGTGTTGGCAACATACCATCTTCA
ACCTCTGCATTGAAAGTTCCCCAATTGAAAGTTGCAGAATCTGCCCAGAGTCCAGCTGCTGCTCATACTACTGATACTGCTGGGTATAATGCAATGGAAGAAAGTGTGAGCAGG
```

GAGAAGCCAGAATTGACAGCTTCAACAGAAAGGGTCAACAAAAGAATGTCCATGGTGGTGTCTGGCCTGACCCCAGAAGAATTTATGCTCGTGTACAAGTTTGCCAGAAAACAC
CACATCACTTTAACTAATCTAATTACTGAAGAGACTACTCATGTTGTTATGAAAACAGATGCTGAGTTTGTGTGTGAACGGACACTGAAATATTTTCTAGGAATTGCGGGAGGA
AAATGGGTAGTTAGCTATTTCTGGGTGACCCAGTCTATTAAAGAAAGAAAAATGCTGAATGAGCATGATTTTGAAGTCAGAGGAGATGTGGTCAATGGAAGAAACCACCAAGGT
CCAAAGCGAGCAAGAGAATCCCAGGACAGAAAGATCTTCAGGGGGCTAGAAATCTGTTGCTATGGGCCCTTCACCAACATGCCCACAGATCAACTGGAATGGATGGTACAGCTG
TGTGGTGCTTCTGTGGTGAAGGAGCTTTCATCATTCACCCTTGGCACAGGTGTCCACCCAATTGTGGTTGTGCAGCCAGATGCCTGGACAGAGGACAATGGCTTCCATGCAATT
GGGCAGATGTGTGAGGCACCTGTGGTGACCCGAGAGTGGGTGTTGGACAGTGTAGCACTCTACCAGTGCCAGGAGCTGGACACCTACCTGATACCCCAGATCCCCCACAGCCAC
TACTGACTGCAGCCAGCCACAGGTACAGAGCCACAGGACCCCAAGAATGAGCTTACAAAGTGGCCTTTCCAGGCCCTGGGAGCTCCTCTCACTCTTCAGTCCTTCTACTGTCCT
GGCTACTAAATATTTTATGTACATCAGCCTGAAAAGGACTTCTGGCTATGCAAGGGTCCCTTAAAGATTTTCTGCTTGAAGTCTCCCTTGGAAATCTGCCATGAGCACAAAATT
ATGGTAATTTTTCACCTGAGAAGATTTTAAAACCATTTAAACGCCACCAATTGAGCAAGATGCTGATTCATTATTTATCAGCCCTATTCTTTCTATTCAGGCTGTTGTTGGCTT

AGGGCTGGAAGCACAGAGTGGCTTGGCCTCAAGAGAATAGCTGGTTTCCCTAAGTTTACTTCTCTAAAACCCTGTGTTCACAAAGGCAGAGAGTCAGACCCTTCAATGGAAGGA
GAGTGCTTGGGATCGATTATGTGACTTAAAGTCAGAATAGTCCTTGGGCAGTTCTCAAATGTTGGAGTGGAACATTGGGGAGGAAATTCTGAGGCAGGTATTAGAAATGAAAAG
GAAACTTGAAACCTGGGCATGGTGGCTCACGCCTGTAATCCCAGCACTTTGGGAGGCCAAGGTGGGCAGATCACTGGAGGTCAGGAGTTCGAAACCAGCCTGGCCAACATGGTG
AAACCCCATCTCTACTAAAAATACAGAAATTAGCCGGTCATGGTGGTGGACACCTGTAATCCCAGCTACTCAGGTGGCTAAGGCAGGAGAATCACTTCAGCCCGGGAGGTGGAG
GTTGCAGTGAGCCAAGATCATACCACGGCACTCCAGCCTGGGTGACAGTGAGACTGTGGCTCAAAAAAAAAAAAAAAAAAAGGAAAATGAAACTAGAAGAGATTTCTAAAAGTC
TGAGATATATTTGCTAGATTTCTAAAGAATGTGTTCTAAAACAGCAGAAGATTTTCAAGAACCGGTTTCCAAAGACAGTCTTCTAATTCCTCATTAGTAATAAGTAAAATGTTT
ATTGTTGTAGCTCTGGTATATAATCCATTCCTCTTAAAATATAAGACCTCTGGCATGAATATTTCATATCTATAAAATGACAGATCCCACCAGGAAGGAAGCTGTTGCTTTCTT
TGAGGTGATTTTTTTCCTTTGCTCCCTGTTGCTGAAACCATACAGCTTCATAAATAATTTTGCTTGCTGAAGGAAGAAAAAGTGTTTTTCATAAACCCATTATCCAGGACTGTT
TATAGCTGTTGGAAGGACTAGGTCTTCCCTAGCCCCCCAGTGTGCAAGGGCAGTGAAGACTTGATTGTACAAAATACGTTTTGTAAATGTTGTGCTGTTAACACTGCAAATAA
ACTTGGTAGCAAACACTTCCAAAAAAAAAAAAAAA

# Mus musculus mRNA for 53BP1 protein

GenBank: AJ414734.1

```
>gi|16754835|emb|AJ414734.1| Mus musculus mRNA for 53BP1 protein
ATGGACCCTACTGGAAGTCAATTGGATTCAGATTTCTCTCAGCAAGACACTCCTTGCCTGATAATAGAAGAT
TCTCAGCCTGAAAGCCAGGTTCTGGAAGAAGATGCAGGCTCTCACTTCAGCGTGCTATCTCGACACCTTCCT
AATCTGCAGATGCACAAAGAGAACCCCGTGTTGGATATTGTATCAAATCCGGAACAATCTGCTGTAGAACAA
GGAGACAGTAATAGCTCATTCAATGAACATCTGAAAGAAAAGAAAGCTTCAGATCCTGTGGAGTCTTCTCAT
TTGGGTACCAGTGGTTCCATCAGTCAGGTCATTGAACGGTTACCTCAGCCAAACAGGACAAGCAGTGCTCTG
GCAGTGACAGTAGAAGCTGCTTCTCTCCCAGAGGAGGAGAAGGAAGAAGAGGAGTTAGAGGAGAAGGAAGGG
GTGGGAGCAAACGCTCCCGGTGCTGACTCCCTTGCTGCTGAAGATTCTGCTTCATCACAGTTGGGCTTTGGA
GTTCTGGAACTGTCCCAGAGCCAGGATGTTGAAGAACATACAGTGCCATATGATGTCAACCAAGAGCATCTG
CAGTTGGTGACCACTAACTCGGGTTCTAGCCCGCTATCTGATGTGGATGCGAGCACTGCAATTAAATGTGAA
GAACAGCCCACTGAAGATATTGCCATGATAGAACAGCCCAGCAAAGACATCCCTGTTACAGTACAGCCCGGT
AAAGGTATCCATGTGGTAGAAGAACAAAATCTACCACTTGTAAGGTCCGAAGACCGGCCGTCCAGTCCTCAA
GTTTCTGTTGCTGCTGTGGAAACAAAGGAACAGGTACCTGCCCGGGAGCTGCTGGAAGAGGGGCCGCAGGTT
CAGCCGTCATCAGAGCCTGAGGTTTCCTCAACCCAGGAGGACTTGTTTGACCAGAGTAGTAAAACAGCTTCT
GATGGTTGTTCTACTCCTTCAAGGGAGGAAGGTGGGTGCTCTCCGGTTTCCACACCTGCTACCACCCTGCAG
CTCCTGCAGCTCTCTGGTCAGAAGCCCCTTGTTCAGGAGAGTCTTTCCACGAATTCCTCAGATCTTGTTGCT
CCTTCCCCTGATGCTTTCCGATCTACCCCTTTTATCGTTCCTAGCAGTCCCACAGAGCAAGGAGGGAGAAAA
GATGAGCCCATGGATATGTCAGTGATACCTGCAGGAGGGGAGCCTTTCCAGAAGCTTCATGATGACGAAGCA
ATGGAGACAGAAAAACCCCTTCTCCCGTCTCAGCCTACTGTGTCACCGCAAGCATCAACACCAGTGTCTCGG
AGCACGCCAGTTTTCACTCCTGGCTCTCTTCCCATCCCGTCCCAGCCCGAGTTCTCTCATGACATTTTCATT
CCATCACCAAGTCTGGAAGAACCATCAGATGATGTGAAGAAAGGTGGAGGTTTACATAGCTCATCTCTTACT
GTTGAGTGTTCCAAGACTTCAGAGAGTGAACCAAAGAATTTCACTGATGACCTTGGGCTCTCCATGACAGGG
GATTCTTGCAAACTGATGCTTTCTACAAGTGAGTATAGTCAGTCCTCAAAGATGGAGAGCTTGGGTTCTCCC
AGGACTGAGGAAGACAGAGAGAACACACAGATTGACGATACTGAACCTTTGTCTCCAGTTAGCAATTCTAAA
CTTCCTGCTGACAGTGAGAATGTCCTGGTGACTCCATCGCAGGACGACCAGGTAGAAATGAGTCAGAATGTA
GATAAAGCAAAAGAGGATGAAACGGAGGACAGAGGTGACTGTAAAGGCAGAGAAGACGCGGTTGCTGAAGAT
GTTTGCATCGACCTCACTTGTGATTCTGGGAGTCAGGCAGTTCCTTCTCCAGCTACCCGTTCGGAGGCACTT
TCCAGTGTCTTAGATCAGGAGGAAGCTATGGACACTAAAGAACACCATCCAGAGGAAGGGTTTTCGGGATCT
GAAGTAGAAGAAGTCCCTGAGACTCCCTGTGGAAGTCACAGAGAGGAGCCCAAAGAAGAAGCGATGGAAAGT
ATCCCACTGCACCTTTCTCTGACTGAAACCCAGTCTGAGGCATTGTGTCTGCAGAAGGAAGCCCCCAAAGAG
GAATGCCCAGAAGCTATGGAAGTTGAAACCAGTGTGATTAGTATTGACTCCCCCCAGAAACTGCAAGTACTT
GACCAGGAATTAGAGCATAAGGACCCAGACACCTGGGAAGAAGCTACTTCTGAGGACTCCAGTGTTGTTATC
GTTGATGTGAAGGAGCCCTCTCCTCGAGCTGATGTTTCCTGTGAACCTTTGGAGGAGGTGGAGAAATGCTCT
GACTCCCAGTCCTGGGAAGGTGTGGCTCCAGAGGAAGAACCATGTGCTGAGAATAGATTAGATACTCCAGAA
GAAAAGCGTATAGAATGTGATGGAGATTCAAAAGCAGAGACCACAGAAAAGGACGCTGTAACAGAAGACTCT
CCACAACCTCCTTTGCCTTCAGTGAGAGACGAACCTGTCAGACCCGATCAGGAGACACAGCAGCCCCAAGTT
CAAGAGAAAGAGAGCCCAGTGACTGTAGATGCAGAAGTGGCTGATGACAAGCAGCTGGGACCAGAGGGCGCA
TGCCAGCAGCTTGAGAAGGCCCCTGCCTGTGCCTCACAGAGTTTCTGTGAAAGTTCTAGTGAAACTCCATTT
CATTTCACTTTGCCTAAAGAAGGTGATATTATCCCACCATTGACTGGCGCAACCCCACCTCTTATTGGGCAC
CTAAAATTGGAGCCCAAGAGACATAGTACTCCTATTGGGATTAGCAACTATCCAGAAAGCACCATAGCAACC
AGTGATGTCACATCTGAAAGCATGGTGGAGATCAATGATCCTCTACTTGGGAATGAAAAAGGGGATTCTGAG
TCTGCCCCAGAAATGGATGGAAAACTGTCTCTGAAAATGAAACTGGTTAGTCCTGAGACAGAGGCCAGT[…]
```

**Figure 2. mRNA sequences of human myosin, human BRCA1 (breast cancer type 1) and murine 53BP1 from NCBI [4] [5] [6].** If the nucleotide sequences evolved to minimize the probability of frameshift mutations we should find AAGAAC and AAGAAT (green) more than AAAAAC and AAAAAT (yellow). The vertical lines reflect the partition into amino acids to highlight that we do not consider for example all AAAAAC sequences but only the ones corresponding to KN, excluding for example --A AAA AC- which is inevitable despite shifts.